\documentstyle[12pt,epsf,aps]{revtex}

\def\prl#1#2#3{{ Phys. Rev. Lett.} {\bf #1}, #2 (#3)}

\def\pla#1#2#3{Phys. Lett. A {\bf #1}, #2 (#3)}
\def\pra#1#2#3{Phys. Rev. A {\bf #1}, #2 (#3)}
\def\prb#1#2#3{Phys. Rev. B {\bf #1}, #2 (#3)}
\def\pre#1#2#3{Phys. Rev. E {\bf #1}, #2 (#3)}

\def\physa#1#2#3{Physica A {\bf #1}, #2 (#3)}
\def\physd#1#2#3{Physica D {\bf #1}, #2 (#3)}

\def\rmp#1#2#3{Rev. Mod. Phys. {\bf #1}, #2 (#3)}

\def\siam#1#2#3{SIAM J. Appl. Math {\bf #1}, #2 (#3)}
\def\ptpo#1#2#3{Prog. Theor. Phys. {\bf #1}, #2 (#3)}

\def\beq{\begin{equation}}
\def\bc{\begin{center}}
\def\ec{\end{center}}
\def\eqn{\end{equation}}

\begin{document}

\title{
Nonequilibrium Dynamics of the Complex Ginzburg-Landau Equation.
I. Analytical Results}

\author{Subir K. Das$^1$, Sanjay Puri$^1$ and M.C. Cross$^2$} 
\address{$^1$School of Physical Sciences, Jawaharlal Nehru University \\
New Delhi -- 110067, India. \\
$^2$Department of Physics, California Institute of Technology \\
Pasadena, California 91125, U.S.A.\\}
\maketitle
\vspace{4cm}
\begin{abstract}
We present a detailed analytical and numerical study of 
nonequilibrium dynamics for the complex Ginzburg-Landau (CGL) equation.
In particular, we characterize evolution morphologies using spiral
defects. This paper (referred to as $\rm I$) is the first 
in a two-stage exposition. Here, we present analytical 
results for the correlation function arising from a single-spiral 
morphology. We also critically examine the utility of the Gaussian
auxiliary field (GAF) ansatz in characterizing a
multi-spiral morphology. In the next paper of this
exposition (referred to as $\rm II$), we will present 
detailed numerical results.
\end{abstract}

\newpage

\section{Introduction}

There have been many studies of pattern formation in the complex 
Ginzburg-Landau (CGL) equation, which has the general form:
\begin{equation}
\label{cgl}
\frac{\partial\psi(\vec r,t)}{\partial t}=\psi(\vec r,t)+
(1+i\alpha)\nabla^2\psi(\vec r,t)-(1+i\beta)|\psi(\vec r,t)|^2\psi(\vec
r,t).
\end{equation} 
In Eq. (\ref{cgl}), $\psi(\vec r,t)$ is a complex order-parameter field
which depends on space ($\vec r$) and time ($t$); 
and $\alpha$, $\beta$ are real parameters. 
The CGL equation arises in diverse contexts, e.g., chemical
oscillations \cite{bz}; thermal convection in binary fluids
\cite{therm}; multi-mode lasers \cite{las}; etc. An overview of
applications of the CGL equation is provided in the review article by
Cross and Hohenberg \cite{hoh}. The importance of the CGL equation
stems from the fact that it provides a generic description of the slow
modulation of oscillations in a spatially-extended system near a Hopf
bifurcation \cite{hopf}.

The CGL equation exhibits a rich range of dynamical behavior with
variation of the parameters $\alpha$ and $\beta$, and the ``phase
diagram'' has been investigated (mostly numerically) by various authors
\cite{rev}. In a large range of parameter space, the emergence and
interaction of spiral (and antispiral) defects plays 
an important role in determining
the morphology. Our present work focuses on characterizing pattern
formation in the CGL equation using spiral-defect structures.

We have analytically and numerically studied 
nonequilibrium dynamics in the CGL equation with
$\alpha=0$. Typically, we consider the evolution morphology
resulting from a small-amplitude random initial
condition. There has been intense research interest in such problems in
the context of far-from-equilibrium statistical physics -- for reviews,
see \cite{revbin,revbray}. The simplest problem in this class considers a
homogeneous two-phase mixture, which has been rendered
thermodynamically unstable by a rapid quench below the critical
co-existence temperature. An example of such a system is a
2-state ferromagnet (in zero magnetic field) at high temperatures, 
which consists of a homogeneous mixture of ``up'' and
``down'' spins. However, below the critical temperature, 
the system prefers to
be in a spontaneously-magnetized state. 
The evolution of the system from the unstable initial
state is a complex nonlinear
process. In appropriate dimensionless units, this evolution is
described by the time-dependent Ginzburg-Landau (TDGL) equation, i.e.,
Eq. (\ref{cgl}) with $\psi(\vec r,t)$ real, and $\alpha=
\beta=0$. The system evolves by the formation and growth of domains
which are enriched in either up or down spins, and are characterized by
a time-dependent length scale, $L(t)$. In the case of a pure and
isotropic ferromagnet, the domain growth law is $L(t)\sim t^{1/2}$, which is
referred to as the Lifshitz-Cahn-Allen (LCA) law \cite{lca}. The
primary mechanism for domain coarsening (or ``phase ordering
dynamics'') is the curvature-driven motion
and annihilation of interfaces (or defects). Ohta {\it et al.}
\cite{ojk} have
formulated an interface-dynamics approach to obtain an analytic form
for the equal-time correlation function of a phase-ordering
ferromagnet. 

Next, let us consider the dynamical XY model, which is Eq. (\ref{cgl})
with $\psi(\vec r,t)$ complex, but $\alpha=\beta=0$. 
In this case, the relevant defect structures (for dimensionality $d\ge
2$) are vortices (or vortex lines, etc.), and domain growth is driven 
by the motion and annihilation of
vortices and anti-vortices. Puri \cite{p} 
has obtained the time-dependent correlation
function for the XY model, using singular-perturbation methods due to
Suzuki \cite{suz}, Kawasaki {\it et al.} \cite{kaw},
and Puri and Roland \cite{pr}. Furthermore, Bray and Puri \cite{bra}
and (independently) Toyoki \cite{toy} have obtained the time-dependent
correlation function for the vector TDGL equation with $O(n)$ symmetry in
$d$ dimensions when $n\leq d$, i.e., when topological defects are
present. (The dynamical XY model corresponds to the case with $O(2)$
symmetry.) The corresponding domain growth law 
is again the LCA law, $L(t)\sim t^{1/2}$, with logarithmic
corrections when $n=d$ \cite{revbray,rb}. 
To the best of our knowledge, there are
no general results available for the case with $n>d$, where the absence of
topological defects makes it difficult to characterize the dynamical
evolution. 

The present two-stage exposition 
focuses on phase ordering dynamics in the CGL
equation with $\alpha=0$. Furthermore, the analytical and numerical
results presented here are for the $2$-dimensional case, where
spirals are point defects. However, the analytical results obtained by
us can be easily extended to the case with $\alpha\neq 0$ and $d\geq 2$,
as the underlying paradigm remains the same, i.e., spiral defects still
determine the morphology in large regions of parameter space and
for higher dimensionality.

Following the work of Hagan \cite{hag}, Aranson {\it et al.} 
\cite{aakw}, and Chate and Manneville \cite{rev}, we briefly
discuss the phase diagram of the $d=2$ CGL equation with $\alpha=0$.
The limit $\beta=0$ corresponds to the dynamical XY model, which is
well understood \cite{p,revbray}. Without loss of generality, 
we consider the case with $\beta\geq 0$. For $0\leq\beta\leq\beta_{1}$ 
($\beta_{1}\simeq 1.397$ \cite{hag}),
spirals (which are asymptotically plane-waves) are linearly stable to
fluctuations. For $\beta_{1} < \beta\leq\beta_{2}$ ($\beta_{2}\simeq
1.82$ \cite{aakw,rev}), spirals are linearly unstable to 
fluctuations, but the growing
fluctuations are advected away, i.e., the spiral structure is globally
stable. Finally, for $\beta_{2}<\beta$, the spirals are globally unstable
structures and cannot exist for extended times \cite{aakw}. Our results
correspond to the parameter regime with $\beta\leq\beta_{2}$.

In this paper (referred to as $\rm I$), we present analytical studies of
the correlation function resulting from single-spiral and 
multi-spiral morphologies. The next paper
(referred to as $\rm II$) presents detailed numerical results and
compares them with the analytical results of $\rm I$.
This paper is organized as follows. In Section $\rm II$, we obtain
analytical results for the correlation function of a 
single-spiral morphology. In Section $\rm III$, we critically
examine the utility
of the Gaussian auxiliary field (GAF) ansatz \cite{revbray} for
the characterization of a multi-spiral morphology. Section $\rm IV$
concludes this paper with a brief summary and discussion of our
analytical results. 

\section{Correlation Function for a Single-Spiral Morphology}

Figure 1 shows a typical evolution from a small-amplitude random initial
condition for the $d=2$ CGL equation with $\alpha=0$ and $\beta=1$.
We have plotted constant-phase regions in this figure, and it is clear
that the evolving morphology is characterized by spirals and their
interactions. (We use the term ``spiral'' for both spirals and
antispirals, unless specifically stated otherwise.) 
There is a characteristic length scale, e.g., 
inter-spiral spacing or square root of inverse defect density, 
which we denote as $L$. Details of our simulation
techniques and comprehensive numerical results will be provided
in paper $\rm II$. Figure 1 is shown here only to motivate our
subsequent discussion.

We would like to quantitatively characterize
the evolution morphology shown in Figure $1$. The standard tool for 
this is the correlation function of the 
order-parameter field \cite{revbin,revbray}, which we will 
define shortly. (The momentum-space structure factor is obtained
as the Fourier transform of the real-space correlation function.) At
the simplest level of approximation, the morphology in the frames of
Figure $1$ can be interpreted as consisting of disjoint spirals, each
of size $L$. (Of course, this overlooks modulations of the 
order-parameter field at spiral-spiral boundaries, but we will discuss
that later.) Therefore, it is obviously of relevance to compute 
the correlation function for a single-spiral solution. 

The CGL equation with $\alpha=0$ has been studied by Hagan \cite{hag}, 
who found that there is a family
of spiral solutions with the following functional form (in $d=2$):
\begin{equation}
\label{hgs}
\psi(\vec r,t)=\rho(r)\exp\left[-i\beta(1-q^2)t+im\theta-i\phi(r)\right] ,
\end{equation}
where $\vec r\equiv (r,\theta)$; $q\geq0$ is a 
constant which is determined uniquely as a function of $\beta$;
and $m$ is the number of arms in the spiral. The cases with $m>0$ and
$m<0$ correspond to a spiral and antispiral, respectively.
The limiting forms of the functions $\rho(r)$ and $\phi(r)$ are
\begin{eqnarray}
\label{rp}
& &\rho(r)\rightarrow (1-q^{2})^{1/2}, ~~~\phi^{\prime}(r)\rightarrow q, 
\hspace{0.5cm}\mbox {as} \hspace{0.5cm} r\rightarrow \infty, \nonumber\\
& &\rho(r)\rightarrow ar^{m}, ~~~\phi^{\prime}(r)\rightarrow r, 
\hspace{0.5cm}~~~~~~~~~\mbox{as}\hspace{0.5cm}r\rightarrow 0,
\end{eqnarray}
where the constant $a$ is determined by finiteness conditions. 
Hagan has presented explicit solutions
for $q(\beta)$ in the cases with $m=1,2$. 
We will focus on the case with $m=\pm 1$, as only
the 1-armed spirals are expected to be stable in the evolution \cite{hag}.
Figure 2 plots Hagan's solution for $q(\beta)$ (with $\beta \leq 1.5$)
in the case with $m= \pm 1$. In the simple limit $\beta=0$, we have $q=0$,
and the spiral solution simplifies to the vortex solution --
for the $m=\pm 1$ vortex, the lines of constant phase correspond to
constant $\theta$. Spiral solutions for the general case with
$\alpha, \beta \neq 0$ have been discussed by Aranson {\it et al.} 
\cite{aakw,akw}.

We are interested in the correlation function for a $1$-armed spiral
at large length scales, so we simplify Eq. (\ref{hgs}) as 
\begin{equation}
\label{hgss}
\psi(\vec r,t)\simeq 
\sqrt{1-q^2}\exp\left[-i\beta(1-q^2)t+i(\theta-qr)\right],
\end{equation}
where we have specialized to the $m=1$ case.
The correlation between points $\vec r_{1}$ and $\vec r_{2}$
is determined as
\begin{eqnarray}
\label{cdef}
C(\vec r_{1},\vec r_{2},t)&=&\mbox{Re}\left\{
\psi(\vec r_{1},t) \psi^{*}(\vec r_{2},t)\right\} \nonumber\\ 
&\simeq&(1-q^2)\mbox{Re}\left\{ \exp [i(\theta_{1}-qr_{1})-
i(\theta_{2}-qr_{2})] \right\} \nonumber \\
& \equiv & C(\vec r_{1}, \vec r_{2}) .
\end{eqnarray}
The average correlation function is obtained by integrating over the
point $\vec r_{1}$, setting $\vec r_{2}=\vec r_{1}+\vec r_{12}$, i.e.,
\begin{eqnarray}
\label{x}
C(r_{12})&=&\frac{1}{V}\int d\vec r_{1}C(\vec r_{1}, \vec r_{1} + \vec r_{12})
h(L-|\vec r_{1}+\vec r_{12}|) \nonumber\\
&=& \frac{(1-q^2)}{V} \mbox{Re} \int d \vec
r_{1} \exp [i(\theta_{1}-\theta_{2}-qr_{1}+q|\vec r_{1}+\vec r_{12}|)]
h(L-|\vec r_{1}+\vec r_{12}|),
\end{eqnarray}
where $V$ is the spiral volume. In Eq. (\ref{x}), we use 
the step function, $h(x)=1~(0)$ if $x \geq 0$ ($x<0$), which ensures 
that we do not include points which lie outside the defect of size $L$.

For $d=2$, the vector notation $\vec r_{2}=\vec r_{1}+\vec r_{12}$ 
is equivalent to $r_{2}e^{i\theta_{2}}=r_{1}e^{i\theta_{1}}+
r_{12}e^{i\theta_{12}}$. Thus, we have
\begin{equation}
e^{i\theta_{2}}=\frac{r_{1}e^{i\theta_{1}}+r_{12}e^{i\theta_{12}}}
{\left[ r_{1}^{2}+{r_{12}}^{2}+2r_{1}r_{12} \cos(\theta_{1}-\theta_{12})
\right]^{1/2}},
\end{equation}
and
\begin{eqnarray}
C(r_{12}) & = & \frac{(1-q^2)}{V} \mbox{Re}
\int_{0}^{L} dr_{1} r_{1} \int_{0}^{2\pi} d\theta_{1}
\frac{r_{1}+r_{12} e^{i(\theta_{1}-\theta_{12})}}
{\left[ r_{1}^{2}+{r_{12}}^{2}+2r_{1} r_{12} \cos(\theta_{1}-
\theta_{12}) \right] ^{1/2}} \times \nonumber\\
&   & \exp\left(-iq\left\{ r_{1}-\left[r_{1}^{2}+r_{12}^{2}+2r_{1}r_{12}
\cos(\theta_{1}-\theta_{12})\right]^{1/2}\right\}\right)
h(L-|\vec r_{1}+\vec r_{12}|).
\end{eqnarray}
We introduce the variables $\theta_{1}-\theta_{12} = \theta$; $x=r_{1}/L$;
$r=r_{12}/L$, to obtain
\begin{eqnarray}
\label{cr}
C(r_{12})=\frac{(1-q^2)}{\pi}\mbox{Re}\int_{0}^{1}dx x\int_{0}^{2\pi}d\theta
\frac{x+re^{i\theta}}
{(x^{2}+r^{2}+2xr\cos\theta)^{1/2}}\times\nonumber\\
\exp \left[ -iqL\left\{ x-(x^{2}+r^{2}+2xr\cos\theta)^{1/2} \right\} \right]
h[1-(x^2+r^2+2xr \cos\theta)^{1/2}],
\end{eqnarray}
where we have used $V=\pi L^2$ in $d=2$.
Thus, the scaling form of the single-spiral correlation function is
$C(r_{12})/C(0)\equiv g(r_{12}/L,q^{2}L^{2})$. In general, 
there is no scaling with the
spiral size because of the additional factor $qL$. We recover scaling
only in the limit $q=0$ ($\beta=0$), which corresponds to the case of a
vortex. Essentially, spirals of different sizes are not
morphologically equivalent because there is more rotation in the phase
as one goes out further from the core. 

Figure 3 plots $C(r_{12})/C(0)$ vs. $r_{12}/L$ for
the case with $\beta=1$ ($q\simeq 0.306$).
These results are obtained by a direct numerical integration of
Eq. (\ref{cr}). We consider $4$ different values of $L$. The functional
form in Figure $3$ exhibits near-monotonic behavior for small values
of $L$ (i.e., in the vortex limit); and pronounced oscillatory 
behavior for larger
value of $L$, as is expected from the integral expression. Notice that
$r_{12}/L\leq 2$ -- larger values of $r_{12}$ correspond to the point
$\vec r_{2}$ lying outside the defect.

Before we proceed, we should point out that the imaginary part of the 
integral in Eq. (\ref{cr}) is non-zero, in general -- corresponding to a weak
correlation between the real and imaginary parts of the order-parameter
field. The imaginary part can also be obtained with relative ease.
However, we will confine our discussion to the conventional
definition of the correlation function in Eq. (\ref{cdef}). Let us next
consider the asymptotic behavior of the correlation function in the limit
$r_{12}/L\rightarrow 0$, though $r_{12}$ is still much larger than 
the size of the defect core $\xi$.

\subsection{Case with $\beta=0$}

In the case with $\beta=0$, we have $q=0$ and the integral expression
in Eq. (\ref{cr}) simplifies as
\begin{equation}
\label{q0}
C(r_{12})=\frac{1}{\pi}\mbox{Re}\int_{0}^{1}dx x\int_{0}^{2\pi}d\theta
\frac{x+re^{i\theta}}
{(x^{2}+r^{2}+2xr\cos\theta)^{1/2}}
h\left[1-(x^{2}+r^{2}+2xr\cos\theta)^{1/2}\right] .
\end{equation}
The behavior in the $r\rightarrow 0$ limit is of considerable interest
as it determines the large-wavevector ($k\rightarrow\infty$) behavior
of the structure factor \cite{revbray}. 
In that case, we can neglect the step function on the right-hand-side
(RHS) of Eq. (\ref{q0}) as it only
provides corrections at the edge of the vortex defect.
Then, after some algebra, we obtain the result 
\begin{equation}
\label{bh}
C(r_{12})=\frac{1}{\pi}\sum_{n=0}^{\infty}\frac{\Gamma(n+\frac{1}{2})^2}
{n!^2}[A_{n}(r)-B_{n}(r)],
\end{equation}
where
\begin{eqnarray}
\begin{array}{r}
{\displaystyle A_{n}(r)=\left \{ \begin{array}{lr}
\frac{2}{5}r^{2}-2r^{2}\ln r, \;~~~~~~~~~~~~~~~~ n=1 & \hfil\\
\\
\frac{(4n+1)}{(n-1)(2n+3)}r^{2}-\frac{1}{(n-1)}r^{2n},\;
~~~n \neq 1, & \\
\end{array}
\right.
}
\end{array}
\end{eqnarray}
and
\begin{eqnarray}
\begin{array}{r}
{\displaystyle B_{n}(r)=\left \{ \begin{array}{lr}
\frac{1}{3}r^{2}-r^{2}\ln  r, \;~~~~~~~~~~~~~~~~~~~~~~~~ n=0  & \hfil\\
\\
\frac{(2n+1)(4n+3)}{2n(n+1)(2n+3)}r^{2}-\frac{(2n+1)}{2n(n+1)}r^{2n+2},\;
~~~n \neq 0. & \\
\end{array}
\right.
}
\end{array}
\end{eqnarray}
This result is implicit in an earlier work of Bray and Humayun
\cite{bh}, who focused upon the singular part of this function.
In the limit $r\rightarrow 0$, the singular terms in $C(r)$
arise from $A_{1}(r)$ and $B_{0}(r)$, and can be computed as
\begin{equation}
C_{\mbox{\scriptsize sing}}(r_{12})=\frac{1}{2}r^{2}\ln  r,
\end{equation}
which gives rise to a power-law tail in the structure factor,
$S(k)\simeq 4\pi L^{2}(kL)^{-4}$, a result referred to as the
``generalized Porod law'' \cite{por,bra}.

\subsection{Case with $\beta \neq 0$}

We would like to undertake a similar asymptotic analysis in the general
case with $\beta\neq 0$. As we are only interested in the limit
$r \rightarrow 0$, we again discard the step function 
on the RHS of Eq. (\ref{cr}). In that case, we obtain
\begin{equation}
\label{crr}
C(r_{12})=\frac{(1-q^2)}{\pi}\mbox{Re}\sum_{n=0}^{\infty}\frac{(iqL)^{n}}{n!}
\int_{0}^{1}dx x e^{-iqLx}\int_{0}^{2\pi}d\theta(x+r\cos\theta)
\left(x^{2}+r^{2}+2xr\cos\theta\right)^{\frac{n-1}{2}}.
\end{equation}
We will separately consider the cases with $n$ odd and $n$ even.\\
(i) \underline{$n$ odd}:

We designate $n=2p+1$ and consider the angular 
integral on the RHS of Eq. (\ref{crr}):
\begin{eqnarray}
\tilde{I}_{2p+1} (x,r)&=&
\int_{0}^{2\pi}d\theta(x+r\cos\theta)(x^{2}+r^{2}+2xr\cos\theta)^{p}
\nonumber\\
&=& 2xr^{2p}\int_{0}^{\pi}d\theta\left(1+\frac{x^{2}}{r^{2}}+
\frac{2x}{r}\cos\theta\right)^{p}+
2r^{2p+1}\int_{0}^{\pi}d\theta \cos\theta \left(1+\frac{x^{2}}{r^{2}}+
\frac{2x}{r}\cos\theta\right)^{p} \nonumber\\
&\equiv& 2xr^{2p}I_{1}+2r^{2p+1}I_{2}.
\end{eqnarray}
The integrals $I_{1}$ and $I_{2}$ are obtained from Gradshteyn and
Ryzhik \cite{gr}, and the consolidated result is
\begin{equation}
\tilde{I}_{2p+1} (x,r)=2\pi\left[\sum_{k=0}^{p}(_{k}^{p})^{2}x^{2k+1}
r^{2(p-k)}+
r^{2}\sum_{k=0}^{\left[\frac{p-1}{2}\right]}(_{k}^{p})(_{k+1}^{p+k})
x^{2k+1}r^{2k}(x^{2}+r^{2})^{p-2k-1}\right],
\end{equation}
where $\left[y\right]$ refers to the integer part of $y$.
The corresponding contribution to $C(r_{12})$ is
\begin{equation}
C_1 (r_{12})=\frac{1-q^{2}}{\pi}\sum_{n=1,3,5,..}^{\infty}
(-1)^{\frac{n-1}{2}}\frac{(qL)^{n}}{n!}
\int_{0}^{1}dx x \sin(qLx) \tilde{I}_n (x,r) .
\end{equation}

The important feature here is that the above expression for 
$C_1 (r_{12})$ only contains powers of $r^{2}$. Therefore, 
the overall contribution to $C(r_{12})$ from this set of terms is analytic
as $r\rightarrow 0$. In the limiting case $q=0$ ($\beta=0$), the above
contribution is identically $0$.\\
(ii) \underline{$n$ even}: 

Next, let us consider the case with $n$ even. We designate $n=2p$, and 
the angular integral on the RHS of Eq. (\ref{crr}) is
\begin{equation}
\tilde{I}_{2p} (x,r)=\int_{0}^{2\pi}d\theta(x+r\cos\theta)
\left(x^{2}+r^{2}+2xr\cos\theta\right)^{p-1/2} .
\end{equation}
Introduce $\rho_{<} = \mbox{min} (x,r)$ and $\rho_{>} = \mbox{max} (x,r)$ 
to obtain
\begin{eqnarray}
\tilde{I}_{2p} (x,r)&=&2\rho_{>}^{2p-1}\int_{0}^{\pi}d\theta(x+r\cos\theta)
\left(1+\frac{\rho_{<}^{2}}
{\rho_{>}^{2}}+\frac{2\rho_{<}}{\rho_{>}}\cos\theta
\right)^{p-1/2} \nonumber\\
&\equiv& 2\rho_{>}^{2p-1}(xI_{3}+rI_{4}).
\end{eqnarray}
The integrals $I_{3}$ and $I_{4}$ can be computed in terms of
hypergeometric functions as follows \cite{gr}:
\begin{equation}
I_{3}=\pi F\left(\frac{1}{2}-p,\frac{1}{2}-p;1;\frac{\rho_{<}^{2}}
{\rho_{>}^{2}}\right),
\end{equation}
and 
\begin{eqnarray}
\label{ir}
I_{4}&=&\pi \left(\frac{1}{2}+p\right) \frac{\rho_{<}}{\rho_{>}}
F\left(\frac{1}{2}-p,\frac{1}{2}-p;2;\frac{\rho_{<}^{2}}{\rho_{>}^{2}}
\right)-\pi \frac{\rho_{<}}{\rho_{>}} F\left(
\frac{1}{2}-p,\frac{1}{2}-p;1;
\frac{\rho_{<}^{2}}{\rho_{>}^{2}}\right)\nonumber\\
&=&\pi\left(p-\frac{1}{2}\right)\frac{\rho_{<}}{\rho_{>}}
F\left(\frac{3}{2}-p,\frac{1}{2}-p;2;
\frac{\rho_{<}^{2}}{\rho_{>}^{2}}\right).
\end{eqnarray}
We have simplified Eq. (\ref{ir}) using the standard identity \cite{as}
\begin{equation}
(c-a-1)F(a,b;c;z)+aF(a+1,b;c;z)=(c-1)F(a,b;c-1;z),
\end{equation}
with $a=1/2-p$, $b=1/2-p$, $c=2$.

Combining the expressions for $I_{3}$ and $I_{4}$, we obtain 
\begin{equation}
\tilde{I}_{2p} (x,r)=2\pi\rho_{>}^{2p-1}
\left[x F\left(\frac{1}{2}-p,\frac{1}{2}-p;1;
\frac{\rho_{<}^{2}}{\rho_{>}^{2}}\right)+r \frac{\rho_{<}}{\rho_{>}}
\left(p-\frac{1}{2}\right)F\left(\frac{3}{2}-p,\frac{1}{2}-p;2;
\frac{\rho_{<}^{2}}{\rho_{>}^{2}}\right)\right].
\end{equation}
The corresponding terms in the correlation function are
\begin{eqnarray}
C_{2}(r_{12})&=& (1-q^{2})\sum_{n=0,2,4,..}^{\infty}(-1)^{n/2}
\frac{(qL)^{n}}{n!}\int_{0}^{1}dx x \cos(qLx) \times \nonumber\\
& & \left[2x\rho_{>}^{n-1}F\left(\frac{1-n}{2},\frac{1-n}{2};1;
\frac{\rho_{<}^{2}}
{\rho_{>}^{2}}\right)+r\rho_{>}^{n-2}\rho_{<}(n-1)
F\left(\frac{3-n}{2},\frac{1-n}{2};2;
\frac{\rho_{<}^{2}}{\rho_{>}^{2}}\right)\right]\nonumber\\
& \equiv & (1-q^{2})(T_{1}+T_{2}).
\end{eqnarray}

The singular contributions to $C(r_{12})$ as $r\rightarrow 0$ arise entirely
from $C_{2}(r_{12})$, as $C_{1}(r_{12})$ is analytic in $r$. A considerable
amount of algebra is involved in extracting the singular terms in
$T_{1}$ and $T_{2}$. For the sake of brevity, we will only sketch the broad
features of the calculation here. We have
\begin{equation}
\label{tr}
T_{1}=2\sum_{p=0}^{\infty}(-1)^{p}\frac{(qL)^{2p}}{(2p)!\Gamma\left(
\frac{1}{2}-p\right)^{2}}\sum_{m=0}^{\infty}\frac{\Gamma\left(
\frac{1}{2}-p+m\right)^{2}}
{m!^{2}}\int_{0}^{1}dx\cos(qLx)x^{2}\frac{\rho_{<}^{2m}}{\rho_{>}^{
2(m-p)+1}},
\end{equation}
where we have used the standard expansion for the hypergeometric
function \cite{as}. The integral on the RHS of Eq. (\ref{tr}) 
can be written as
\begin{equation}
\label{is}
I_{5}=\frac{1}{r^{2(m-p)+1}}\int_{0}^{r}dx\cos(qLx)x^{2m+2}+
r^{2m}\int_{r}^{1}dx\cos(qLx)x^{-2(m-p)+1}.
\end{equation}
The first term on the RHS of Eq. (\ref{is}) 
is analytic as $r\rightarrow 0$. The
second term contributes singular terms only if $m\geq p+1$, 
yielding the result
\begin{equation}
I_{5}=(-1)^{m-p}\frac{(qL)^{2(m-p-1)}}{[2(m-p-1)]!}r^{2m}\ln r+
\mbox{analytic terms}.
\end{equation}
Replacing this in the expression for $T_{1}$, some algebra yields
\begin{equation}
T_{1}=\sum_{p=0}^{\infty}\sum_{m=0}^{\infty}(-1)^{p+m+1}
\frac{(qL)^{2(p+m)}}{(2p)!(2m)!}\frac{\Gamma\left(\frac{3}{2}+m\right)^{2}}
{\Gamma\left(\frac{1}{2}-p\right)^{2}(m+p+1)!^{2}}r^{2(m+p+1)}\ln r+
\mbox{analytic terms}.
\end{equation}
A similar analysis for $T_{2}$ yields
\begin{eqnarray}
T_{2}&=&\sum_{p=0}^{\infty}\sum_{m=0}^{\infty}(-1)^{p+m}
\frac{(qL)^{2(p+m)}}{(2p)!(2m)!}\frac{\Gamma\left(\frac{1}{2}+m\right)^{2}}
{\Gamma\left(\frac{1}{2}-p\right)^{2}(m+p)!^{2}}
\frac{(2m+1)}{(m+p+1)}r^{2(m+p+1)}\ln 
r+\nonumber\\
& & ~~~~~~~~~~~~~~~~~~~~~~~~~~~~~~~~~~~~~~~~~~~~~~~~~~~~\mbox{analytic terms}.
\end{eqnarray}

We can combine the singular terms from $T_{1}$ and $T_{2}$ to obtain
the singular part of $C(r_{12})$ as follows:
\begin{eqnarray}
C_{\mbox{\scriptsize sing}}(r_{12})&=&
\frac{1}{2}\sum_{p=0}^{\infty}\sum_{m=0}^{\infty}(-1)^{p+m}
\frac{(qL)^{2(p+m)}}{(2p)!(2m)!}\frac{\Gamma\left(\frac{1}{2}+m\right)^{2}}
{\Gamma\left(\frac{1}{2}-p\right)^{2}(m+p+1)!^{2}}\times\nonumber\\
& & ~~~~~~~~~~~~~~~~~~~(2m+1)(2p+1)r^{2(m+p+1)}\ln r .
\end{eqnarray}
We notice that the leading-order singularity is unchanged and continues
to be $C_{\mbox{\scriptsize sing}}(r_{12}) \simeq
\frac{1}{2}r^{2}\ln  r$, as in the case with $\beta=0$. However,
there is now a sequence of sub-dominant 
singularities proportional to $(qL)^{2}r^{4}\ln r$,
$(qL)^{4}r^{6}\ln r$, etc., and these become increasingly important as the
length scale $L$ increases. These sub-dominant terms in
$C_{\mbox{\scriptsize sing}}(r_{12})$ are reminiscent of the leading-order
singularities in models with $O(n)$ symmetry, where $n$ is even 
\cite{revbray,bh}.
Of course, in the context of $O(n)$ models, these singularities only
arise for $n\leq d$ as there are no topological defects unless this
condition is satisfied. In the present context, all these terms are
already present for $d=2$. The implication for
the structure-factor tail is a sequence of power-law decays with
$S(k)\sim (qL)^{2(m-1)}L^{d}/(kL)^{d+2m}$, where $m=1,2$, etc. 
Thus, though the true asymptotic behavior in $d=2$ 
is still the generalized Porod tail, $S(k) \sim L^2 (kL)^{-4}$, it may be
difficult to disentangle this from other power-law decays.

The results presented in this section are of relevance in determining
the small-distance behavior of the correlation function, or the
large-wavevector behavior of the structure factor. This is because
small length-scales only probe individual defects. Nevertheless,
as our numerical results in paper II will demonstrate, the single-spiral
correlation function agrees with the correlation function for
multi-spiral morphologies (obtained numerically) over a considerable
range of distances. For even larger
length-scales, we have to explicitly account for the modulation of the
order parameter at defect-defect boundaries. We address this problem in
the next section of this paper.

\section{Utility of Gaussian Auxiliary Field Ansatz for a 
Multi-Spiral Morphology} 

The evolution in Figure 1 is characterized by a morphology with
multiple spirals and anti-spirals. Initially, spirals and anti-spirals 
are attracted to each other and annihilate, thereby
decreasing the defect density and increasing the inter-defect
distance (or characteristic length scale). When the defect density
is large, the spiral sizes are small and spirals are similar to
vortices. Therefore, we expect an initial coarsening regime
which is analogous to that for the XY model -- both in terms of the
domain growth law, $L(t) \sim (t/\ln t)^{1/2}$ \cite{par}; and the
morphology, as characterized by the correlation function \cite{p,bra}.
This is in accordance with our numerical simulations, as we discuss in
paper II. Distinctive effects of spirals are seen for length scales
$L > L_c$, where $L_c \sim q^{-1}$ -- clearly, $L_c \rightarrow \infty$
as $q \rightarrow 0$ (or $\beta \rightarrow 0$). Furthermore, there is 
a repulsive spiral-antispiral potential beyond a certain distance, 
which prevents the annealing of all defects \cite{hoh,akw}. Thus, the
evolving system ``freezes'' (in a statistical sense) into a
multi-spiral morphology.
This should be contrasted with the case of the dynamical XY model
($\alpha = \beta = 0$ in Eq. (\ref{cgl})), where we expect the 
zero-temperature system to continue coarsening as $t\rightarrow\infty$.

A common theme in the characterization of dynamical evolution with a
nonconserved order parameter is the introduction of a Gaussian
auxiliary field (GAF) \cite{revbray,ojk,bra,maz}. Essentially, 
the GAF ansatz
takes the form $\psi(\vec r,t)=F[m(\vec r,t)]$, where the function
$F[m]$ is determined from the defect structure, and the complex field $m$
(which measures the location relative to the defect core) is assumed to
obey a Gaussian distribution. The zero-crossings of the field $m$
correspond to the location of defect cores. The GAF ansatz 
enables a straightforward
computation of the correlation function for the field $\psi(\vec r,t)$.
However, the analytical justification for the GAF ansatz is meagre and its
primary virtue appears to be that it works rather well in
some situations \cite{revbray}.

Let us examine the utility of the GAF ansatz in the present context.
The appropriate form of the ansatz for the CGL equation in the regime
where the spiral structures are well-developed
is (using Hagan's solution for the spiral defect)
\begin{equation}
\label{gaf}
\psi(\vec r,t) \simeq \frac{\sqrt{1-q^{2}}m(\vec r,t)}{\sqrt{1-q^{2}+
|m(\vec r,t)|^{2} }}\exp \left[-i(\omega t+q|m(\vec r,t)|) \right] ,
\end{equation}
where $\omega=\beta(1-q^{2})$; and we take $|\psi| \simeq |m|$ 
near the defect core ($|m|\rightarrow 0$), 
in accordance with Hagan's solution. The
field $m (=m_{1}+im_{2})$ is assumed to obey a 
Gaussian distribution with
\begin{equation}
P(m_{i})=\frac{1}{\sqrt{2\pi\sigma^{2}}}\exp\left(
-\frac{m_{i}^{2}}{2\sigma^{2}}\right),~~~i=1,2~~,
\end{equation}
where $\sigma^{2}=\left<m_{i}(\vec r,t)^{2}\right>$; and
the fields $m_{1}(\vec r,t)$ and $m_{2}(\vec r,t)$ are taken to
be statistically independent of each other.

Our numerical results show that the GAF ansatz in Eq. (\ref{gaf}) is
reasonable in the vicinity of defects. However, it is inappropriate for
defect-defect boundaries, where the order-parameter amplitude
$| \psi |$ is often larger than $\sqrt{1-q^2}$. This is demonstrated 
in Figure 4, which replots Figure 1 with defect locations marked 
by asterisks; and regions where $| \psi | > \sqrt{1-q^2}$ marked in 
black. As discussed before, for early times (e.g., $t=25$), the system
evolution is governed by the interaction of vortices. Thus, the
appropriate GAF ansatz should have max($| \psi |) = 1$, as in the
case of the XY model. For late times (e.g., $t=1000$), the system
has well-developed spirals. Nevertheless, the GAF
ansatz for the order-parameter field is obviously
inappropriate for large regions of space at these parameter values.
For other values of $\beta$, the same general arguments apply though
there are changes in the crossover time to spiral-mediated growth;
and the fraction of spatial region where the GAF ansatz is unreasonable.

Let us examine the validity of the GAF
ansatz in regions where $| \psi | < \sqrt{1-q^2}$. We
can simplify the ansatz in Eq. (\ref{gaf}) by defining
the variable $m^{\prime}=me^{-i\phi}$, where 
$\phi=\omega t+q|m|$. Then, we have the
corresponding probability distribution for (say) $m_{1}^{\prime}$ as
\begin{eqnarray}
P^{\prime}(m_{1}^{\prime})&=&\int_{-\infty}^{\infty}
dm_{1}\int_{-\infty}^{\infty}dm_{2}\delta(m_{1}^{\prime}-m_{1}\cos\phi-m_{2}
\sin\phi)P(m_{1})P(m_{2})\nonumber\\
&=&\frac{1}{2\pi\sigma^{2}}\int_{-\infty}^{\infty}
dm_{1}\int_{-\infty}^{\infty}dm_{2}\delta(m_{1}^{\prime}-m_{1}\cos\phi-m_{2}
\sin\phi)\exp \left(-\frac{m_{1}^{2}+m_{2}^{2}}{2\sigma^{2}}\right).
\end{eqnarray}
As usual, we transform ($m_{1},m_{2}$) $\rightarrow$ ($|m|,\theta$) to
obtain 
\begin{equation}
P^{\prime}(m_{1}^{\prime})=\frac{1}{2\pi\sigma^{2}}\int_{0}^{\infty}d|m|
|m|\exp \left(-\frac{|m|^{2}}{2\sigma^{2}}\right)
\int_{0}^{2\pi}d\theta\delta(m_{1}^{\prime}-|m|\cos(\theta-\phi)).
\end{equation}
Because of the periodicity of the function $\cos(\theta-\phi)$, 
the phase factor $\phi$ is inconsequential and 
\begin{equation}
\label{gf}
P^{\prime}(m_{1}^{\prime}) = 
\frac{1}{\sqrt{2\pi\sigma^{2}}}\exp\left(-\frac
{{m_{1}^{\prime}}^{2}}{2\sigma^{2}}\right) ,
\end{equation}
and a similar distribution also applies for the variable $m_{2}^{\prime}$.

Thus, we have the appropriate GAF ansatz (dropping primes) as follows:
\begin{equation}
\label{sg}
\psi(\vec r,t)=\frac{\sqrt{1-q^{2}} m(\vec r,t)}{\sqrt{1-q^{2} + |m(\vec
r,t)|^{2}}} ,
\end{equation}
where the variables $m_{1}(\vec r,t)$ and $m_{2}(\vec r,t)$ ($m(\vec
r,t)= m_{1}(\vec r,t)+i m_{2}(\vec r,t)$) are Gaussian and independent
of each other. The inverse relation between the
variables $\psi$ and $m$ is
\begin{equation}
\label{fm}
m(\vec r,t)=\frac{\sqrt{1-q^{2}} \psi(\vec r,t)}
{\sqrt{1-q^{2}-|\psi(\vec r,t)|^{2}}}.
\end{equation}

We want to examine the validity of the GAF ansatz numerically 
\cite{yos,dr} in the context of the evolution depicted in Figure 1
(or Figure 4). The appropriate parameter values are $\beta = 1$ and
$q \simeq 0.306$ \cite{hag} (see Figure 2). In Figure 5, we plot
the single-variable distribution for the field $m_{1}(\vec r,t)$, 
obtained directly from our simulation of the CGL equation using 
Eq. (\ref{fm}) in regions where $| \psi | < \sqrt{1-q^2}$. 
The data in Figure 5 is obtained as
an average over 5 independent runs for $N^2$-lattices, with $N=512$.
(Details of our simulation will be provided in paper $\rm II$ of this
exposition.) Figure 5(a) is a plot of $P(m_1)$ vs. $m_1$ from 4
different times -- corresponding to the evolution pictures shown
in Figure 1. In Figure 5(b), we have scaled variables and superposed
the data for $P(m_{1}) \sigma$ vs. $m_{1}/\sigma$, where $\sigma$
is obtained from the best-fit of the numerical data to the functional
form in Eq. (\ref{gf}). The data collapses onto a single master curve, 
which is reasonably approximated by the Gaussian form
$P(x)=\frac{1}{\sqrt{2\pi}}e^{-\frac{x^{2}}{2}}$,
denoted as a solid line in Figure 5(b).

Figure 5 has been obtained by focusing only on regions where
$| \psi | < \sqrt{1-q^2}$, which is essentially equivalent to
considering disjoint spirals, for which the correlation function
has already been obtained in Section II. We have examined various
ad-hoc methods of improving the GAF ansatz in Eq. (\ref{sg}). For
example, one could set the saturation amplitude of the order
parameter to its maximum value ($| \psi |_{\mbox{\scriptsize sat}} 
\simeq 1$ for Figure 1), rather than $| \psi |_{\mbox{\scriptsize sat}} 
= \sqrt{1-q^2}$. Figure 6 plots the resultant probability distributions,
$P(m_1)$ vs. $m_1$, with $| \psi |_{\mbox{\scriptsize sat}} = 1$.
For early times ($t=25$), the distribution has a Gaussian form, as
expected from our analogy with the XY model. However, with the
emergence of well-formed spirals, the distribution develops a double-peak
and is clearly non-Gaussian.

We have also studied some other possible ways of rectifying the
GAF ansatz. We find that these ad-hoc approaches invariably
result in non-Gaussian distributions for the auxiliary field. Perhaps
a more honest approach should be based on the order-parameter field
for spiral-spiral pairs \cite{akw} as a function of two independent
auxiliary fields -- referring to distances from the centers of the
two spirals. We are presently studying the utility of such an
approach for characterizing the multi-spiral morphology.

\section{Summary and Discussion}

Let us conclude this paper with a brief summary and discussion of our
results. We have undertaken a detailed analytical and numerical
investigation of nonequilibrium dynamics in a special case of the
complex Ginzburg-Landau (CGL) equation. Our results are described in a
two-stage exposition. This paper (referred to as $\rm I$) 
constitutes the first stage of this exposition, 
and describes analytical results for the
time-dependent correlation function. Our analytical arguments rely on
the significance of spiral-defect structures in
determining the morphology and evolution of the CGL equation from a
random initial condition.

In this paper, we describe results for 
the exact correlation function $C(r_{12})$ of a single spiral
defect of size $L$, and undertake its asymptotic analysis in the limit
$r_{12}/L \rightarrow 0$ but $r_{12}/\xi \gg 1$, where $\xi$ is the
size of the defect core. We find that there is a sequence of
singularities in this limit, which are reminiscent of singularities for
defects with $O(n)$ symmetry, where $n$ is even. However, the dominant
singularity as $r_{12}/L \rightarrow 0$ corresponds to the case of
vortex defects, as expected.
The implications for the large-wavevector tail of the
structure factor are also discussed. 

We also investigate the validity of the Gaussian auxiliary field 
(GAF) ansatz in the context of multi-spiral morphologies. For early
times ($L < L_c \sim q^{-1}$), domain growth in the CGL equation is
analogous to that for the XY model, whose domain growth law and
correlation function are well understood \cite{revbray}. For later
times, we find that the simple
GAF ansatz is not reasonable, as it is unable to account for order-parameter
modulations in the defect-defect boundaries. We have attempted ad-hoc
improvements of the GAF ansatz but these invariably result in non-Gaussian
distributions for the corresponding auxiliary field. We are presently
investigating the possibility of formulating a generalized GAF ansatz
in terms of the order-parameter field for a spiral-spiral pair.

More generally, the utility of the GAF ansatz arises from the
summation over phases from many defects, which results in a near-Gaussian
distribution for the auxiliary field. However, in the present context, 
the shocks between spirals effectively isolate one spiral
region from the influence of other regions. As a matter of fact, the
waves from other spirals decay exponentially through the shock and the
phase of a point is always dominated by the nearest spiral. Therefore,
we expect that the correlation function will be dominated by the
single-spiral result -- in accordance with our numerical results.

The next paper in this exposition
(referred to as $\rm II$) will present detailed
numerical results for phase ordering dynamics in the CGL equation. 
In particular, we will focus upon the crossover from vortex-mediated
dynamics (at early times) to spiral-mediated dynamics (at late times).
Furthermore, we will compare our numerical results 
for the correlation function of the order-parameter field with the
analytic form for a single-spiral defect presented in this paper.

Before we conclude this paper, it is worth stressing 
that the results presented are
easily adaptable to the general case of the CGL equation
with $\alpha,\beta \neq 0$. Again, the evolving morphology in a large region
of parameter space is characterized by the presence and annihilation of
spirals and anti-spirals \cite{rev}. 
The results of the present paper apply
directly in that case also, with minor modifications in the functional
forms of the spiral solution in Section $\rm II$.

\section*{Acknowledgements}

SP is grateful to A.J. Bray and H. Chate for useful discussions. SKD
is grateful to the University Grants Commission, India, for financial
support in the form of a research fellowship.

\newpage

\section*{Figure Captions}

\vskip0.5cm
\noindent{\bf Figure 1:} Evolution of a small-amplitude random initial
condition for the complex Ginzburg-Landau (CGL) equation with $\alpha=0$,
$\beta=1$. These evolution pictures were obtained from an isotropic
Euler discretization of Eq. (\ref{cgl}), implemented on an 
$N^{2}$-lattice ($N=256$) with periodic boundary conditions 
in both directions. The
discretization mesh sizes were $\Delta t=0.01$ and $\Delta x=1.0$. The
pictures show regions of constant phase $\theta_{\psi}=\tan^{-1}
(\mbox{Im}\psi/\mbox{Re}\psi)$, measured in radians, with the
following coding: $\theta_{\psi} \in [1.85, 2.15]$ (black); 
$\theta_{\psi} \in [3.85, 4.15]$ (red); $\theta_{\psi} \in [5.85, 6.15]$ 
(green). The snapshots are labeled by the appropriate evolution times. \\
\ \\
{\bf Figure 2:} Plot of $q(\beta)$ vs. $\beta$ for the $1$-armed
spiral solution of the CGL equation with $\alpha=0$. (cf.
Figure 5 of Ref. \cite{hag}.) \\
\ \\
{\bf Figure 3:} Correlation function for the $1$-armed spiral
solution when $\beta=1$ ($q \simeq 0.306$). We plot $C(r_{12})/C(0)$ vs.
$r_{12}/L$ for different spiral sizes, $L=10,25,50,100$ -- denoted by the
specified line-types. The results are obtained from a direct numerical
integration of Eq. (\ref{cr}). \\
\ \\
{\bf Figure 4:} Evolution shown in Figure 1, replotted to clarify 
the utility of the GAF ansatz in this context. The asterisks 
denote spiral centers; and regions where $| \psi | > \sqrt{1-q^2}$ 
are shaded black. \\
\ \\
{\bf Figure 5:} (a) Plot of data for $P\left[m_{1}(\vec r,t)\right]$ 
vs. $m_{1}(\vec r,t)$ from $4$ different times, $t=25,50,100,1000$ -- 
denoted by the symbols
shown. The parameter values are identical to those in Figure $1$. We
use Eq. (\ref{fm}) to obtain data for $m_{1}(\vec r,t)$ directly from
the order-parameter field in our numerical solution of the CGL
equation -- considering only regions where $| \psi | < \sqrt{1-q^2}$. 
The data was obtained as an average over 5
independent runs for $N^{2}$-lattices ($N=512$). \\
(b) Scaled plot of data from Figure 5(a). We superpose data for
$P\left[m_{1}(\vec r,t)\right] \sigma (t)$ vs. 
$m_{1}(\vec r,t)/\sigma (t)$, where $\sigma (t)$ is obtained from the
best-fit of the numerical data to a Gaussian distribution.
The solid line refers to the Gaussian function 
$P(x)=\frac{1}{\sqrt{2\pi}}e^{-x^{2}/2}$.  \\
\ \\
{\bf Figure 6:} Plot of data for $P\left[m_{1}(\vec r,t)\right]$ 
vs. $m_{1}(\vec r,t)$ from times $t=25,50,100,1000$ -- denoted 
by the symbols shown. The parameter values and statistical details
are identical to those for Figure 5(a). Data for $m_{1}(\vec r,t)$
is obtained directly from the numerical data for $\psi (\vec r,t)$,
using Eq. (\ref{sg}) with $| \psi |_{\mbox{\scriptsize sat}} =
\sqrt{1-q^2}$ replaced by $| \psi |_{\mbox{\scriptsize sat}} = 1$.

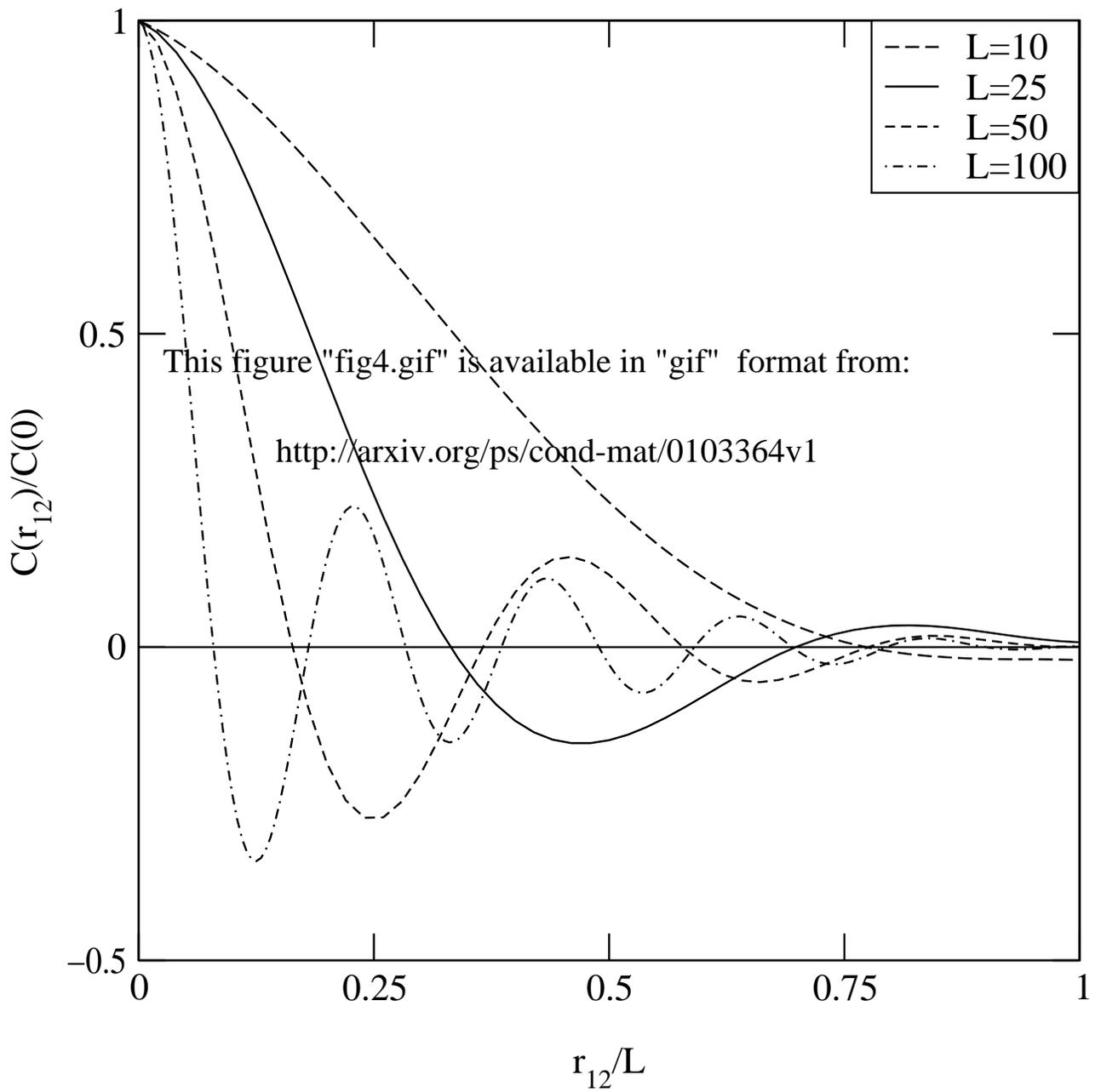
\end{document}